\documentclass[aps,prb,floatfix,showpacs,preprint,amsmath,amssymb,]{revtex4-1}

\usepackage{graphicx}
\usepackage{dcolumn}
\usepackage{bm}
\usepackage{color}
 \usepackage{url}

\begin{document}


\title{Micromagnetic approach to exchange bias---A new look at an old problem}

\author{J. Dubowik}
\affiliation{Institute of Molecular Physics, Polish Academy of Physics, M. Smoluchowskiego 17, PL-60-179, Pozna{\'n}, Poland}
\author{I. Go{\'s}cia{\'n}ska}
\affiliation{Faculty of Physics, A. Mickiewicz University, Umultowska 85, PL-61-614, Pozna{\'n}, Poland}

\date{\today}

\begin{abstract}
We present a micromagnetic approach to {the} exchange bias (EB) in ferromagnetic (FM)/antiferromagnetic (AFM) thin film systems with a small number of irreversible interfacial magnetic moments.  We express the exchange bias field $H_{EB}$  in terms of {the} fundamental micromagnetic length scale of FM---the exchange length $l_{ex}$. The benefit from this approach is a better separation of {the} factor related to {the} FM layer from {the} factor related  to {the} FM/AFM coupling at interfaces. Using this approach we estimate  the {upper} limit of $H_{EB}$ in real FM/AFM systems.
\end{abstract}


\maketitle

\section{Introduction \label{Sec:1}}
\begin{figure}
\includegraphics{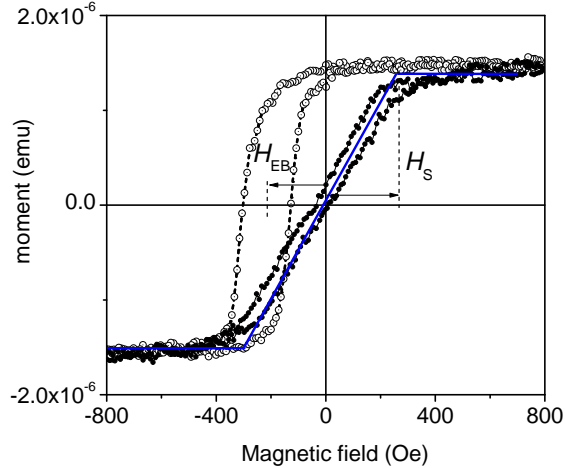}
\caption{\label{fig:1} Magnetization $(M)$ versus magnetic
field $(H)$ curves of a sample with the following layer structure:
Si (substrate)/IrMn(20 nm)/Co(4 nm)/IrMn(20 nm). Dashed {line} and open circles: field parallel to the exchange
bias field. Full circles and {blue} lines: field perpendicular to the exchange bias field. The film was prepared under the same conditions as described in Ref.~\onlinecite{dubowik}}
\end{figure}

The coupling between a ferromagnet (FM) and  an antiferromagnet (AFM) that is  set up on field cooling from temperatures above the N\'{e}el temperature of the AFM results in an exchange bias (EB). \cite{bean} However, it seems  that we  do not yet have   a general and compact micromagnetic description  of EB in spite of a number of numerical simulations \cite{suss, schuller, lambertoduo} and models. \cite{nogues,berkowitz}  In this respect, three main points need to be emphasized.
(\textit{\textbf{i}}) In numerous {proposed} mesoscopic and microscopic models of EB \cite{nogues, berkowitz}, the master formula for the unidirectional anisotropy field $H_{EB}$ (the exchange bias field) is
\begin{equation}\label{eq:1}
 H_{EB} = \frac{J_{EB}}{ M\,\, t_{FM}},
\end{equation}
where   $J_{EB}$ is the interfacial exchange bias energy  and $t_{FM}$ is the thickness of {the} FM layer with magnetization $M$. Equation {(}\ref{eq:1}{)} represents a micromagnetic relation expressing the equilibrium between the   exchange bias energy density $J_{EB}/t_{FM}$ and the Zeeman energy. \cite{bean}  The main problem in this relation is $J_{EB}$, which is generally ill{-}defined{,} so that we do not know how it is determined by the fundamental parameters of a ferromagnet, taking into account the peculiarities of the interface structure, and of the antiferromagnet. For example, if we arbitrarily suppose that $J_{EB}$ is determined exclusively by the AFM, then, in accordance with Eq.~(\ref{eq:1}), $H_{EB}$ should somehow decrease for EB systems with FM of a high magnetization (e.g., Co).  However, we will show in Sec. \ref{Sec:3} that the tendency is in general quite opposite. A ferromagnet with a high magnetization and a high exchange stiffness gives usually the highest $H_{EB}$.
(\textit{\textbf{ii}}) An important step forward in explaining {the} magnitude of the EB has been done by St\"{o}ehr's group,  who, using x-ray magnetic circular dichroism, showed that EB is produced by a small  ($\approx0.04=4$\%) number of irreversible AFM spins.\cite{stohr, stohrbook} Therefore, a spin structure at an FM/AFM interface consists of two groups. {First, t}he uncompensated AFM spins---weakly coupled to the rest of the AFM spin lattice so that they can rotate. Second, the  irreversible spins---a small fraction of uncompensated spins that are tightly coupled  to the AFM spin lattice.  Hence, a reduction factor $\epsilon$ equal to the fraction of irreversible spins should be taken into account if Eq.~(\ref{eq:1}) is to explain the experimental data.
(\textit{\textbf{iii}}) In FM/AFM bilayers, the coercive field $H_{C}$ of the FM undergoes a substantial increase {by a factor} of 10--20 in comparison {with} a single FM film due to an anisotropy ${\langle} K {\rangle}$  imposed on the FM by the AFM's uncompensated spins. \cite{nogues} However, after inspecting  a large number of available experimental data, \cite{kohn, harres} it appears that the saturation field $H_{S}$ (measured in the hard direction) of the FM coupled to an AFM  is {a}  more reliable quantity than $H_{C}$ measured in {the} easy direction. Figure \ref{fig:1} serves as a typical example showing that $H_{S}$ is of the order of $H_{EB}$. Therefore, the uniaxial  anisotropy field $H_{S}$  is
\begin{equation}\label{eq:2}
 H_{S} = \frac{2\,\, {\langle}K {\rangle}}{M}.
\end{equation}

The aim of the present paper is to  express Eq.~{(}\ref{eq:1}{)} in a more fundamental form involving the  micromagnetic characteristics  of the FM. It should be stressed that by using   magnetic measurements of  FM/AFM systems with EB, we can only determine the magnetic properties of the FM. Nevertheless, looking from the FM side, we can argue about an interaction between the FM and the AFM. We shall concentrate on the most important aspects of EB, namely the main factors that determine the values of $H_{EB}$ and $J_{EB}$. In particular, we shall determine the role of the FM layer and we shall look for an FM/AFM system that fulfils the requirements of an almost ideal EB effect.
\section{Model \label{Sec:2}}
Three  quantities describe the micromagnetism of ferromagnets: the magnetization $M$, the magnetic anisotropy $K$, and the exchange stiffness constant $A$. Since the exchange interactions are relevant for a long range spin ordering, we postulate that the last quantity is mainly responsible for interactions between the FM and AFM layers. In the first approximation{,} let {us} assume that an FM/AFM interface is ideal, so that the interfacial spins are fully coupled.  It can be easily derived from {the} definition of the exchange interaction in the Heisenberg approach  that  $J_{EB}$ can be approximated by $2\,{\langle} A{\rangle}/\xi$, where ${\langle} A{\rangle}$ is the average exchange stiffness of an FM layer within an interface region with a thickness $\xi$ of the order of the lattice parameter.\cite{suss}   Hence Eq.~{(}\ref{eq:1}{) takes the form}
\begin{equation}\label{eq:1a}
 H_{EB} = \frac{2 {\langle} A{\rangle}}{ M\,\,\xi\, t_{FM}}.
\end{equation}

 {Equivalently}, the micromagnetic characteristics of an FM can be expressed in terms of  {the} exchange length $l_{ex}$ and the exchange correlation length $l_{cor}$ (domain wall parameter) defined as
\begin{eqnarray}\label{eq:1b}
l_{ex}=\sqrt{({A/2\pi M^{2}})}, \\ \nonumber \\
  \nonumber l_{cor}=\sqrt{({A/K})},
\end{eqnarray}
respectively. Both  $l_{ex}$ and  $l_{cor}$  are the fundamental length scales  that control the behavior of magnetic material{s} and are relevant  for {the} description of an inhomogeneous orientation of the spin structure. \cite{skomski} In Tab.~\ref{tab:table1}{,} we gathered  the values of the magnetic polarization $4\pi M$ and the exchange stiffness $A$ necessary for {the} estimation of $l_{ex}$ for the typical soft magnetic materials,  some Heusler alloys, and magnetite.   The values of $l_{ex}$ are within the range of $3 - 8 \times 10^{-7}$ cm = 3--8 nm, while $l_{cor}$  (of {the} order of the Bloch wall thickness) {varies} considerably from $l_{cor}\approx1$  nm in hard magnetic materials to over 100 nm in soft ferromagnets.\cite{skomski} The spin-wave stiffness $D=2g \mu_{B} A/M$ is also included for comparison{,} since both $A$ and $D$ are frequently  used to describe the stiffness of exchange interactions. $g$ is {the} Land{\'{e}} g-factor and $\mu_{B}$ is the Bohr magneton.

\begin{figure}
\includegraphics{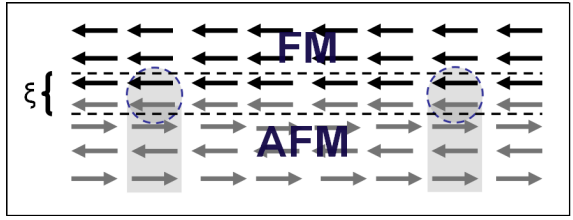}
\caption{\label{fig:2}Schematic diagram of the ferromagnet/an\-ti\-fer\-ro\-mag\-net (FM/AFM) interface of thickness $\xi$ with uncompensated spins and  the spins (marked by dashed  circles) pinned to the AFM spin lattice (shaded area).}
\end{figure}

By multiplying and dividing Eq.~{(}\ref{eq:1a}{)} by $4\pi M$, we can express it in a different way:
\begin{equation}\label{eq:3}
 H_{EB} = 4 \pi M \, \frac{{\langle} A{\rangle}}{2\pi\,M^2} \, \frac{1}{\xi \, t_{FM}}=4 \pi M \,\frac{ {\langle} l_{ex}{\rangle}^{2}} {\xi \, t_{FM}},
\end{equation}
where  ${\langle} l_{ex}{\rangle}$ denotes an averaged exchange length within the interface region. For a typical value of  $l_{ex}=5$ nm (see Tab.~\ref{tab:table1}), Eq.~ {(}\ref{eq:3}{)} {leads to} an unrealistically high value of $H_{EB}\approx 170$ kOe if we assume typical  values for $4\pi M\approx10$ kG, $\xi\approx 3\times 10^{-8}$ cm, and $t_{FM}=l_{ex}$.
\begin{table}
\caption{\label{tab:table1}Basic magnetic parameters of ferromagnetic $\textit{3-d}$ metals, typical soft magnetic alloys, some Heusler alloys and magnetite (Fe$_3$O$_4$): the demagnetizing field $4 \pi M$, the exchange stiffness constant $A$, the spin-wave stiffness $D$, the exchange length $l_{ex}$, and the product of demagnetizing field and the square of exchange length. }
\begin{ruledtabular}
\begin{tabular}{cccccc}
Material & $4\pi M$ & $A$ & $D$ & $l_{ex}$ & $4\pi M~l_{ex}^2$ \\
&  (kOe) & ($\mu$erg/cm) & (meV nm$^2$)  & (nm)  & ($10^{9}$Oe cm$^2$)\\
\hline
Fe& 21.4\footnotemark[1] &2.0\footnotemark[2] & 2.8& 3.3 & 2.3
 \\
Co& 18.1\footnotemark[1] & 2.5\footnotemark[2] & 4.5 & 4.4 & 3.5
\\
Ni& 6.1\footnotemark[1] & 0.8\footnotemark[2] & 4.5 & 7.6 & 3.5
 \\
Ni$_{80}$Fe$_{20}$& 10.0\footnotemark[1]& 1.0\footnotemark[2] & 2.5  & 5.0 & 2.5
 \\
Co$_{47}$Fe$_{53}$& 17.0\footnotemark[3] & 5.9\footnotemark[3] & 8.0  & 7.2 & 8.7
\\
Co$_2$FeSi& 14.1\footnotemark[4]  & 3.2\footnotemark[4] & 7.0 & 6.4 & 5.7
\\
Co$_2$MnSn& 9.9 \footnotemark[5]& 0.6\footnotemark[5] & 2.0 & 3.9 & 1.5
 \\
Ni$_2$MnSn& 5.1 \footnotemark[6]& 0.1\footnotemark[6] &0.4 & 2.6 & 0.3
 \\
Fe$_3$O$_4$ \footnotemark[8]& 5.9 \footnotemark[7]& 0.7\footnotemark[7] &5.0 & 7.1 & 3.0
 \\

\end{tabular}
\end{ruledtabular}
\footnotetext[1]{from Ref.~\onlinecite{ohandley}.}
\footnotetext[2]{from Ref.~\onlinecite{frait}.}
\footnotetext[3]{from Ref.~\onlinecite{liu}.}
\footnotetext[4]{from Ref.~\onlinecite{gaier}.}
\footnotetext[5]{from Ref.~\onlinecite{yilgin}.}
\footnotetext[6]{from Ref.~\onlinecite{dubowik1}.}
\footnotetext[7]{from Ref.~\onlinecite{zaag}.}
\footnotetext[8] {All data are representative for room temperature except that of magnetite, which is at 5 K.}
\end{table}
The  comment (\textbf{\textit{ii}}) implies  that ${\langle A \rangle}$ and   ${\langle}l_{ex}{\rangle}$  are {to some extent} weakened by {the} low number
of the pinned spins. Let us inspect the impact of {the} low number of the irreversible spins on ${\langle}A{\rangle}$
{more closely}. If we imagine the interface shown in Fig. \ref{fig:2} with the spins (marked by circles)  pinned to the rest of the AFM (marked by shaded area), we can see that they are exchange{-}coupled with equal numbers {$\epsilon$ of} FM spins. Therefore{,}
\begin{equation}\label{eq:4}
 {\langle} A{\rangle}\approx \frac{I\,\,\,
\epsilon \, S_{FM}\,\,\epsilon \,S_{AFM}}{a}\approx \epsilon^2\, A,
\end{equation}
where $I$ is the exchange integral.  $S_{FM}$  {and} $S_{AFM}$ are the FM and AFM spins, respectively. Here we assume that the EB systems exhibit negative bias, so that the FM spins and irreversible AFM spins are aligned in the same direction ($I>0$). \cite{stohrbook}  Hence, an EB for a realistic interface with a low number of irreversible spins {can be} expressed by
\begin{equation}\label{eq:5}
 H_{EB} = 4 \pi M \, l_{ex}^{2}\, \frac{\epsilon^2}{ \xi \, t_{FM}}
\end{equation}
with  the product of $4\pi M\,l_{ex}^{2}$ as the leading {factor}.
From Tab.~\ref{tab:table1} one can see that the leading {factor} is highest for {the} Co-Fe alloy and, unexpectedly, for the Co$_2$FeSi Heusler alloy, while it does not vary much  for \mbox{3-\textit{d}} metals and Ni-Fe.   It is easy to show that an $\epsilon$ on the order of a few percent {provides} a realistic value of $H_{EB}\approx 200$ Oe ($\epsilon=0.015$), as is shown in Fig. \ref{fig:1}, for example,  and in agreement with other experimental data (see, for example Ref.~\onlinecite{nogues} and references therein). The {expression} ${\langle} l_{ex}{\rangle} = \epsilon \,l_{ex}$ can be regarded as an effective exchange length in FM due to the low number of irreversible AFM spins. Note, however, that ${\langle} l_{ex}{\rangle}$ is on the order of only 0.1 nm if $\epsilon = 0.04$. This is a remarkably small value since, by definition, the exchange length is the length below which atomic exchange interactions dominate over typical magnetostatic fields. \cite{skomski}

Accordingly, the interfacial exchange bias energy  $J_{EB}$ is expressed by
\begin{equation}\label{eq:5a}
 J_{EB} = 2 \pi M^2 (  \frac{2 l_{ex}^{2}\,\epsilon^2}{ \xi}),
\end{equation}
where the second factor in parentheses has the dimension of length scale of 200 nm if $\epsilon=1$ so that  $J_{EB}$ would take a huge value of 50--150 erg/cm$^{2}$.  However, since $J_{EB}$ must be less than $K_{AFM}\times t_{AFM}$, \cite{bean} the upper limit of $J_{EB}$ is less than 10 erg/cm$^{2}$ if $K_{AFM}\approx 10^{7}$ erg/cm$^{3}$ and $t_{AFM}\approx 5$ nm (a typical critical value of AFM thickness). \cite{coehoorn}  It is noteworthy that if $\epsilon=0.35$ and 0.04, $J_{EB}$ would be 10--4  and  0.4--0.1 erg/cm$^{2}$, respectively.

\section{ Discussion \label{Sec:3}}
\begin{figure}
\includegraphics{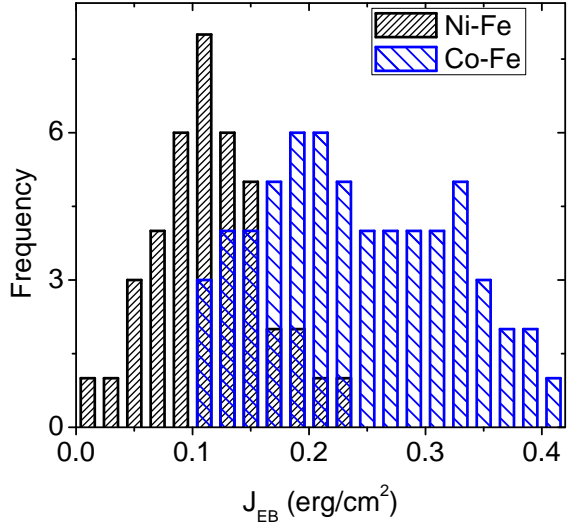}
\caption{\label{fig:2a}Histograms showing distribution of the experimental values of $J_{EB}$ taken from Tab. 13 in Ref.~\onlinecite{coehoorn}. The histograms  show distribution of the data for Ni-Fe (Ni$_{80}$Fe$_{20}$) and for Co-Fe films (Co$_{90}$Fe$_{10}$), respectively. Frequency has the meaning of the number of  data falling in a bin of $J_{EB}$}
\end{figure}

The lowest limit of $J_{EB}$ of 0.1--0.4 erg/cm$^{2}$ is worth comment.
Since research into EB encompasses a huge number of papers, we can draw useful conclusions on the average exchange bias energy  $J_{EB}$ using simple statistics for a large number of experimental data under the assumption that each experiment is of equal significance. A collection of tabulated data gathered by Coehoorn \cite{coehoorn} is an invaluable database. We gathered the distribution of the values of $J_{EB}$ (taken from Tab. 13 in Ref.~\onlinecite{coehoorn}) in the form of histograms as shown in Fig. \ref{fig:2a}. It is clearly seen that both for Ni-Fe and Co-Fe layers in contact with various metallic random substitutional fcc-type AFM alloys, the distributions of  the data have the shape of a normal distribution even though the width of the histogram for the Co-Fe data is four times higher than that of the Ni-Fe data.  Most important, however, is that the mean value of $J_{EB}$ is 0.1  and 0.25 erg/cm$^{2}$ for Ni-Fe and Co-Fe, respectively. If we estimate the values of $J_{EB}$ making use of Eq.~(\ref{eq:5a}) assuming $\epsilon = 0.04$, we unexpectedly arrive at  $J_{EB}=0.11$  and 0.24 erg/cm$^{2}$, nearly the same as the mean values evaluated from the histograms for Ni-Fe and Co-Fe, respectively.
For the calculations, we took the appropriate data from Tab.~\ref{tab:table1} and $\xi=0.3$ and 0.25 nm for permalloy and Co, respectively. The remarkable agreement between the experimental and calculated values of  $J_{EB}$ may seem a coincidence, but it may also suggest that the assumed ratio of the irreversible pinned spins of just a few percent is typical of metallic FM/AFM systems. One of the highest values of $J_{EB}$  ever measured for metallic FM/AFM bilayers (with a high  $4\pi M~l_{ex}^{2}$  factor) is 1.3 erg/cm$^{2}$ for  Co$_{70}$Fe$_{30}$ in contact with a highly L1$_{2}$ ordered Mn$_{3}$Ir phase. \cite{tsunoda} A rough estimate employing Eq.~(\ref{eq:5a}) suggests that such a ``giant'' value is achieved for a merely twofold increase in $\epsilon$ to 0.08--0.09.

Therefore, the FM/AFM systems behave as if a specific localized exchange coupling between $\epsilon$ fraction of FM spins and an equal fraction of AFM irreversible spins  were blurred in a delocalized ``sea'' of FM interacting spins. In this aspect, EB can be regarded as a  perturbation in the exchange energy of the FM  in contact with AFM.

Still, Eq.~(\ref{eq:5})  describes  an  idealized case of the exchange bias.  In reality, in most cases both the FM and the AFM have a large number of defects: they consist of grains {o}n a nanometer scale with grain boundaries, etc. Nevertheless, Eq.~(\ref{eq:5}) captures the most important material and interface characteristics that determine the order of magnitude of  the exchange bias {on the nanoscale}. The most characteristic in Eq.~(\ref{eq:5})  is that the factor $4 \pi M \, l_{ex}^2$ depends exclusively on the FM (due to our assumption that the interface coupling between the irreversible AFM spins and the FM spins is positive), while the factor $\epsilon^2$  depends mostly on the AFM (its anisotropy) and the quality of the interface.

By applying the same transformation to Eq.~(\ref{eq:2}) as to Eq.~{(}\ref{eq:3}{)}, we have
\begin{equation}\label{eq:6}
 H_{S} = 4 \pi M \frac{{\langle} K{\rangle}}{2 \pi M^2}\, \frac{\epsilon^2 A}{\epsilon^2 A}=4 \pi M \, l_{ex}^{2} \, \frac {1}{l_{cor}^2},
\end{equation}
which has the same symmetrical form as Eq.~{(}\ref{eq:5}{)} with $l_{cor}^2$ in the denominator. It is characteristic that the {factor} $\epsilon^2$ is absent. For a typical value of $H_{S}=200$ Oe in Fig. \ref{fig:1}, Eq.~(\ref{eq:6}) leads to $l_{cor}=32$ nm.


\begin{figure}
\includegraphics{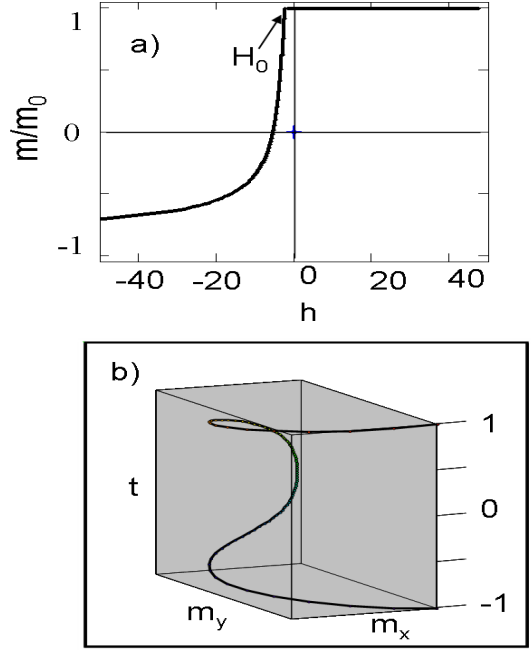}
\caption{\label{fig:3} (a) Asymmetrical dependence of reduced magnetization $m$ versus reduced magnetic field $h$ applied opposite to the pinning direction. (b) Twisting of the magnetization vector
defined as $\vec{m}={\vec{i}}\, cos \Theta(t)+{\vec{j}}\, sin \Theta(t)$ with the distance between two  FM/AFM interfaces for $h=0.5$. The magnetization is fully pinned at $t=\pm 1$.}
\end{figure}

The presence of both $l_{ex}$ and $l_{cor}$  in Eqs {(}\ref{eq:5}{)} and {(}\ref{eq:6}{)} may be linked with an inhomogeneous spin structure of the FM and suggests the formation of a magnetization twisting. Such a magnetization twisting was analyzed about 50 years ago   by Aharoni \textit{et al.} \cite{aharoni}  They considered  a ferromagnetic slab, infinite in the $x$ and $y$ directions and of width $2t_{FM}$. At $t=z/t_{FM}=\pm 1$, the spins are assumed to be held in the $x$ direction by the exchange coupling with the AFM and expressed with appropriate boundary conditions. Let an external field $H$ be applied in the $x$ direction. The easiest mode for magnetization changes is evidently rotation of the spins in the $xy$ plane. The functions which minimize the energy of such a system are the
solutions of {the} Euler equation
\begin{equation}\label{eq:8}
 \frac{d^2 \Theta}{dt^2}-\frac{h}{(t_{FM}/l_{ex})^2} \sin \Theta=0,
\end{equation}
with the boundary conditions $\Theta'(0)=0$, $\Theta(\pm1)=0$ and with $h=H/ 4 \pi M$. The solution of Eq.~{(}\ref{eq:8}{)} is expressed in terms of the complete elliptic integral of the first kind $K_{C}$
\begin{equation}\label{eq:9}
\Theta (t)=2 \arcsin \big( k\; sn [(1-t) K_{C}(k),k]\big),
\end{equation}
where $sn$ is the sine amplitude function and $k$ is ``hidden'' in the relation $-h=K_{C}^2 (k)/(t_{FM}/l_{ex})^2$. This solution leads to a strongly asymmetric magnetization reversal curve as shown in Fig. \ref{fig:3} (a), which saturates at $-h\approx \infty$. This important approach to EB has not attracted much attention except in some old papers. \cite{salansky} Similar asymmetric magnetization reversals have been recently observed  in Ni/FeF$_2$ bilayers and interpreted as originating from the intrinsic broken symmetry of the system, which results in local incomplete domain walls parallel to the interface (i.e., the magnetization twisting) in reversal to negative saturation of the FM. \cite{schuller} The twisting of the magnetization vector shown in Fig. \ref{fig:3} (b) comes from the boundary conditions stating that the magnetization is fully ``free'' at the center of the FM layer and fully ``pinned'' at the two FM/AFM interfaces.

As is seen in Fig. \ref{fig:3} (a), the magnetization  starts twisting above a certain field ${h}>(\pi /2)^2/(t_{FM}/l_{ex})^2$. Hence,
\begin{equation}\label{eq:10}
 H_{0} =4 \pi M\, l_{ex}^2\; \frac{\left(\pi/2\right)^2}{t_{FM}^2}.
\end{equation}
Equation {(}\ref{eq:10}{)} describes the magnitude of  the exchange bias field for an ideal FM/AFM system without uniaxial anisotropy imposed by AFM but with fully irreversible spins at the interfaces.  The similarity between Eq.~{(}\ref{eq:10}{)} and Eq.~{(}\ref{eq:5}{)}  is striking in that the factor $(\pi/2)^{2}/t_{FM}^{2}$ is purely geometrical.
If we equate $H_{EB}$ with $ H_{0}$   (Eq.~(\ref{eq:5}) to Eq.~{(}\ref{eq:10}{)}) in order to estimate the maximal value that $\epsilon$ can achieve for the ideal pinning described by the boundary conditions, we obtain
\begin{equation}\label{eq:11}
\epsilon_{max} ^2 = (\pi/2)^2 \frac{\xi}{t_{FM}}.
\end{equation}
For typical values  $\xi \approx 0.3$ nm and $t_{FM} \approx l_{ex} \approx 5-10$ nm, ${\epsilon}^{2} \approx 0.15 - 0.075$. As a result 38\%--27\% of the irreversible AFM spins would produce the highest possible values that $H_{EB}$ ($J_{EB}$) can achieve, i.e., 25--12~kOe (6--3 erg/cm$^{2}$). Hence, we come to the conclusion that the highest value that ${\epsilon}$ can attain  is (38\%--27\%) is just due to the formation of an incomplete domain wall (i.e., magnetization twisting). In reality, however, AFM is polycrystalline and defected, so that these values are overestimated. \cite{lambertoduo}

It appears that among a large number of FM/AFM systems all-oxide Fe$_3$O$_4$/CoO epitaxial bilayers  nearly satisfy the requirements of ideal pinning with $J_{EB}=2.1$ erg/cm$^{2}$---the value is only about 8 times smaller than the exchange biasing estimated according to the Meiklejohn--Bean model. \cite{zaag,zaagbook} $J_{EB}$ estimated from Eq.~(\ref{eq:5a}) with ${\epsilon}^{2} = 0.13$ (i.e., $\approx1/8$) and $\xi \approx 0.8$ nm (lattice parameter of Fe$_3$O$_4$) yields $J_{EB}=2.04$ erg/cm$^{2}$, which justifies the [100] oriented Fe$_3$O$_4$/CoO bilayer system's being very close to the straightforward idea concerning the micromagnetic approach. It is noteworthy that polarized neutron reflectivity measurements of similar Fe$_3$O$_4$/NiO multilayers have provided evidence of domain-wall formation (magnetization twisting) in the exchange-biased state but within the ferromagnetic, rather than the AFM, layer. \cite{ball} A question may be posed: why is the EB the highest in all-oxide FM/AFM systems? Neglecting the complex interplay between the microstructure of the AFM layer and the FM/AFM interface, it seems that superexchange coupling via the intervening p-orbitals of the oxygen atoms plays a leading role. The coupling, between the magnetic ions with half-occupied orbitals (Fe$^{2+}$ and Co$^{2+}$) through the intermediary oxygen ion, of the superexchange is indirect (the magnetic ions are of 0.4 nm apart) and  strongly antiferromagnetic. What seems even more important, is that both oxide lattices are based on an approximately close-packed lattice of oxygen ions with Co$^{2+}$ and Fe$^{3+}$ in tetrahedral  or octahedral (Fe$^{3+}$ and Fe$^{2+}$) interstitial.  \cite {zaagbook} This feature makes Fe$_3$O$_4$/CoO epitaxial bilayers a model system to study EB. \cite{zaag} In contrast, in all-metallic FM/AFM systems, the exchange coupling between the FM and AFM species is direct, so that any change in ordering  at the interfaces results in a frustration of exchange interactions. \cite{radu}

\begin{figure}
\includegraphics{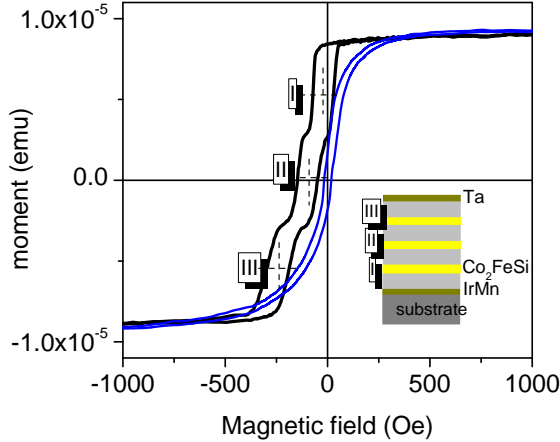}
\caption{\label{fig:5} Hysteresis loops of [Ta 5 nm/(IrMn 20 nm/Co$_2$FeSi 10 nm)$\times 3$/ IrMn 20 nm/ Ta 5 nm] multilayer taken with the magnetic field applied parallel (black) and perpendicular (blue) to $H_{EB}$ direction, respectively. The inset shows the multilayer structure consisting of a stack of  IrMn (gray)/Co$_2$FeSi (yellow)/IrMn three-layers.}
\end{figure}
Now we can understand why in most of the FM/AFM all-metallic thin film systems the EB field is of the  order of  100--400 Oe. As seen in Tab.~\ref{tab:table1}, the product $4 \pi M \,l_{ex}^2$ does not differ much {among} most of the soft FM materials.
Therefore, any enhancement of the EB relies mainly on increasing $\epsilon$.   We have little room for manoeuvre except to increase $\epsilon$ by some technological trick like, for example, dusting the interfaces with ultrathin Co or Mn layers \cite{endo, dubowik} or  a proper setting AFM in a magnetic field. \cite{nogues} Specifically, $\epsilon$ determines the quality of setting AFM on cooling from $T>T_{N}$. \cite{ogrady} A spectacular example of such a gradual improvement in EB is observed in a Ta 5/(IrMn 20/Co$_2$FeSi 10)$\times 3$/ IrMn 20/ Ta 5 multilayer annealed at 400$^o$C for 15 min. and field cooled to room temperature  (Fig. \ref{fig:5}). The details of the sample preparation can be found in Ref.~\onlinecite{dubowik}. As seen in Fig. \ref{fig:5}, the hysteresis loop  (black) taken with the magnetic field parallel to  the exchange bias consists of three loops related to the subsequent Co layers in the stack. These three loops, of equal heights and nearly equal coercive field of $\sim 50$ Oe, are shifted along the field axis by  $H_{EB}$ of 20, 72 and 235 Oe for the first (I), second (II) and third (III) Co layer, respectively. Simultaneously, in accordance with the discussion of Eq.~(\ref{eq:6}), the estimated value of $H_{S}$  is 150 Oe (see Fig. \ref{fig:5}, hysteresis taken with a magnetic field applied perpendicular to $H_{EB}$ direction). It has been confirmed (see Ref.~\onlinecite{dubowik} for details) with a magneto-optical Kerr magnetometer that the upper (III) Co layer has the highest $H_{EB}$. A simple estimate employing Eq.~(\ref{eq:5}) using the data  from Tab.~\ref{tab:table1} for Co$_2$FeSi gives the values of $\epsilon$ of 0.01, 0.02, and 0.05 for the layers I, II, and III, respectively.  As was discussed above, $\epsilon$ depends on the interface quality and on the anisotropy of the AFM layer. Since the interfaces in the  stack should not differ much, the increase in anisotropy seems to be responsible for these slight changes in $\epsilon$. We have proven with x-ray diffraction measurements (see Ref.~\onlinecite{dubowik}) that the grain size of IrMn increases  as the subsequent layers are deposited from the substrate, so that the increase in AFM anisotropy is justified.  However, in view of our discussion, we do not expect that $\epsilon$ can exceed the values of several percent in the case of all-metallic FM/AFM systems.

\section{ Summary\label{Sec:4}}

 In summary, we have shown that the exchange bias resulting from a coupling between FM and AFM layers can be described in terms of a rough micromagnetic approach, which seems to capture the essential characteristics of the exchange bias. Specifically, we showed that  the interfacial interactions involved between the FM and the AM  {reduce} to a geometrical problem  with the fundamental micromagnetic length scale being the exchange length $l_{ex}$. The model identifies the range of the exchange bias field $H_{EB}$ (exchange bias energy $J_{EB}$) compatible with those observed in experiment.  Using the model, we proved that  the highest effective number of irreversible spins is lower than $\sim$30\%--40\%.


\bibliography{micromagnetic}

\begin{thebibliography}{28}%
\makeatletter
\providecommand \@ifxundefined [1]{%
 \@ifx{#1\undefined}
}%
\providecommand \@ifnum [1]{%
 \ifnum #1\expandafter \@firstoftwo
 \else \expandafter \@secondoftwo
 \fi
}%
\providecommand \@ifx [1]{%
 \ifx #1\expandafter \@firstoftwo
 \else \expandafter \@secondoftwo
 \fi
}%
\providecommand \natexlab [1]{#1}%
\providecommand \enquote  [1]{``#1''}%
\providecommand \bibnamefont  [1]{#1}%
\providecommand \bibfnamefont [1]{#1}%
\providecommand \citenamefont [1]{#1}%
\providecommand \href@noop [0]{\@secondoftwo}%
\providecommand \href [0]{\begingroup \@sanitize@url \@href}%
\providecommand \@href[1]{\@@startlink{#1}\@@href}%
\providecommand \@@href[1]{\endgroup#1\@@endlink}%
\providecommand \@sanitize@url [0]{\catcode `\\12\catcode `\$12\catcode
  `\&12\catcode `\#12\catcode `\^12\catcode `\_12\catcode `\%12\relax}%
\providecommand \@@startlink[1]{}%
\providecommand \@@endlink[0]{}%
\providecommand \url  [0]{\begingroup\@sanitize@url \@url }%
\providecommand \@url [1]{\endgroup\@href {#1}{\urlprefix }}%
\providecommand \urlprefix  [0]{URL }%
\providecommand \Eprint [0]{\href }%
\providecommand \doibase [0]{http://dx.doi.org/}%
\providecommand \selectlanguage [0]{\@gobble}%
\providecommand \bibinfo  [0]{\@secondoftwo}%
\providecommand \bibfield  [0]{\@secondoftwo}%
\providecommand \translation [1]{[#1]}%
\providecommand \BibitemOpen [0]{}%
\providecommand \bibitemStop [0]{}%
\providecommand \bibitemNoStop [0]{.\EOS\space}%
\providecommand \EOS [0]{\spacefactor3000\relax}%
\providecommand \BibitemShut  [1]{\csname bibitem#1\endcsname}%
\let\auto@bib@innerbib\@empty
\bibitem [{\citenamefont {Meiklejohn}\ and\ \citenamefont {Bean}(1956)}]{bean}%
  \BibitemOpen
  \bibfield  {author} {\bibinfo {author} {\bibfnamefont {W.~H.}\ \bibnamefont
  {Meiklejohn}}\ and\ \bibinfo {author} {\bibfnamefont {C.~P.}\ \bibnamefont
  {Bean}},\ }\href@noop {} {\bibfield  {journal} {\bibinfo  {journal} {Phys.
  Rev.}\ }\textbf {\bibinfo {volume} {102}},\ \bibinfo {pages} {1413} (\bibinfo
  {year} {1956})}\BibitemShut {NoStop}%
\bibitem [{\citenamefont {S\"{u}ss}(2002)}]{suss}%
  \BibitemOpen
  \bibfield  {author} {\bibinfo {author} {\bibfnamefont {D.}~\bibnamefont
  {S\"{u}ss}},\ }\emph {\bibinfo {title} {Micromagnetic simulations of
  antiferro- and ferromagnetic structures for magnetic recording}},\ \href@noop
  {} {\bibinfo {type} {{Ph.D.} thesis}},\ \bibinfo  {school} {Wien University,
  Austria} (\bibinfo {year} {2002})\BibitemShut {NoStop}%
\bibitem [{\citenamefont {Li}\ \emph {et~al.}(2006)\citenamefont {Li},
  \citenamefont {Petraci\^{c}}, \citenamefont {Morales}, \citenamefont
  {Olamit}, \citenamefont {Batlle}, \citenamefont {Liu},\ and\ \citenamefont
  {Schuller}}]{schuller}%
  \BibitemOpen
  \bibfield  {author} {\bibinfo {author} {\bibfnamefont {Z.-P.}\ \bibnamefont
  {Li}}, \bibinfo {author} {\bibfnamefont {O.}~\bibnamefont {Petraci\^{c}}},
  \bibinfo {author} {\bibfnamefont {R.}~\bibnamefont {Morales}}, \bibinfo
  {author} {\bibfnamefont {J.}~\bibnamefont {Olamit}}, \bibinfo {author}
  {\bibfnamefont {X.}~\bibnamefont {Batlle}}, \bibinfo {author} {\bibfnamefont
  {K.}~\bibnamefont {Liu}}, \ and\ \bibinfo {author} {\bibfnamefont {I.~K.}\
  \bibnamefont {Schuller}},\ }\href@noop {} {\bibfield  {journal} {\bibinfo
  {journal} {Phys. Rev. Lett.}\ }\textbf {\bibinfo {volume} {96}},\ \bibinfo
  {pages} {217205} (\bibinfo {year} {2006})}\BibitemShut {NoStop}%
\bibitem [{\citenamefont {Fecioru-Morariu}\ \emph {et~al.}(2010)\citenamefont
  {Fecioru-Morariu}, \citenamefont {Nowak},\ and\ \citenamefont
  {G\"{u}ntherodt}}]{lambertoduo}%
  \BibitemOpen
  \bibfield  {author} {\bibinfo {author} {\bibfnamefont {M.}~\bibnamefont
  {Fecioru-Morariu}}, \bibinfo {author} {\bibfnamefont {U.}~\bibnamefont
  {Nowak}}, \ and\ \bibinfo {author} {\bibfnamefont {G.}~\bibnamefont
  {G\"{u}ntherodt}},\ }in\ \href {http://dnb.d-nb.de} {\emph {\bibinfo
  {booktitle} {Magnetic properties of antiferromagnetic oxide materials}}},\
  \bibinfo {editor} {edited by\ \bibinfo {editor} {\bibfnamefont
  {L.}~\bibnamefont {D\'{u}o}}, \bibinfo {editor} {\bibfnamefont
  {M.}~\bibnamefont {Finazzi}}, \ and\ \bibinfo {editor} {\bibfnamefont
  {F.}~\bibnamefont {Ciccacci}}}\ (\bibinfo  {publisher} {Wiley-VCH},\ \bibinfo
  {year} {2010})\ Chap.~\bibinfo {chapter} {5}\BibitemShut {NoStop}%
\bibitem [{\citenamefont {Nogues}\ and\ \citenamefont
  {Schuller}(1999)}]{nogues}%
  \BibitemOpen
  \bibfield  {author} {\bibinfo {author} {\bibfnamefont {J.}~\bibnamefont
  {Nogues}}\ and\ \bibinfo {author} {\bibfnamefont {I.~K.}\ \bibnamefont
  {Schuller}},\ }\href@noop {} {\bibfield  {journal} {\bibinfo  {journal} {J.
  Magn. Magn. Mater.}\ }\textbf {\bibinfo {volume} {192}},\ \bibinfo {pages}
  {203} (\bibinfo {year} {1999})}\BibitemShut {NoStop}%
\bibitem [{\citenamefont {{A.E. Berkowitz and K. Takano}}(1999)}]{berkowitz}%
  \BibitemOpen
  \bibfield  {author} {\bibinfo {author} {\bibnamefont {{A.E. Berkowitz and K.
  Takano}}},\ }\href@noop {} {\bibfield  {journal} {\bibinfo  {journal} {J.
  Magn. Magn. Mater.}\ }\textbf {\bibinfo {volume} {200}},\ \bibinfo {pages}
  {552} (\bibinfo {year} {1999})}\BibitemShut {NoStop}%
\bibitem [{\citenamefont {Ohldag}\ \emph {et~al.}(2001)\citenamefont {Ohldag},
  \citenamefont {Regan}, \citenamefont {St\"{o}hr}, \citenamefont {Scholl},
  \citenamefont {Nolting}, \citenamefont {Luening}, \citenamefont {Stamm},
  \citenamefont {Anders},\ and\ \citenamefont {White}}]{stohr}%
  \BibitemOpen
  \bibfield  {author} {\bibinfo {author} {\bibfnamefont {H.}~\bibnamefont
  {Ohldag}}, \bibinfo {author} {\bibfnamefont {T.~J.}\ \bibnamefont {Regan}},
  \bibinfo {author} {\bibfnamefont {J.}~\bibnamefont {St\"{o}hr}}, \bibinfo
  {author} {\bibfnamefont {A.}~\bibnamefont {Scholl}}, \bibinfo {author}
  {\bibfnamefont {F.}~\bibnamefont {Nolting}}, \bibinfo {author} {\bibfnamefont
  {J.}~\bibnamefont {Luening}}, \bibinfo {author} {\bibfnamefont
  {C.}~\bibnamefont {Stamm}}, \bibinfo {author} {\bibfnamefont
  {S.}~\bibnamefont {Anders}}, \ and\ \bibinfo {author} {\bibfnamefont {R.~L.}\
  \bibnamefont {White}},\ }\href@noop {} {\bibfield  {journal} {\bibinfo
  {journal} {Phys. Rev. Lett.}\ }\textbf {\bibinfo {volume} {87}},\ \bibinfo
  {pages} {247201} (\bibinfo {year} {2001})}\BibitemShut {NoStop}%
\bibitem [{\citenamefont {St\"{o}hr}\ and\ \citenamefont
  {Siegmann}(2006)}]{stohrbook}%
  \BibitemOpen
  \bibfield  {author} {\bibinfo {author} {\bibfnamefont {J.}~\bibnamefont
  {St\"{o}hr}}\ and\ \bibinfo {author} {\bibfnamefont {H.~C.}\ \bibnamefont
  {Siegmann}},\ }\href@noop {} {\emph {\bibinfo {title} {Magnetism from
  fundamentals to nanoscale dynamics}}},\ Series: Springer Series in
  Solid-State Sciences, Vol. 152\ (\bibinfo  {publisher} {Springer},\ \bibinfo
  {address} {Berlin},\ \bibinfo {year} {{2006}})\ \bibinfo {type}
  {Section}~\bibinfo {chapter} {10}, pp.\ \bibinfo {pages}
  {100--119}\BibitemShut {NoStop}%
\bibitem [{\citenamefont {Kohn}\ \emph {et~al.}(2011)\citenamefont {Kohn},
  \citenamefont {Dean}, \citenamefont {Kovacs}, \citenamefont {Zeltser},
  \citenamefont {Carey}, \citenamefont {Geiger}, \citenamefont {Hrkac},
  \citenamefont {Schrefl},\ and\ \citenamefont {Allwood}}]{kohn}%
  \BibitemOpen
  \bibfield  {author} {\bibinfo {author} {\bibfnamefont {A.}~\bibnamefont
  {Kohn}}, \bibinfo {author} {\bibfnamefont {J.}~\bibnamefont {Dean}}, \bibinfo
  {author} {\bibfnamefont {A.}~\bibnamefont {Kovacs}}, \bibinfo {author}
  {\bibfnamefont {A.}~\bibnamefont {Zeltser}}, \bibinfo {author} {\bibfnamefont
  {M.~J.}\ \bibnamefont {Carey}}, \bibinfo {author} {\bibfnamefont
  {D.}~\bibnamefont {Geiger}}, \bibinfo {author} {\bibfnamefont
  {G.}~\bibnamefont {Hrkac}}, \bibinfo {author} {\bibfnamefont
  {T.}~\bibnamefont {Schrefl}}, \ and\ \bibinfo {author} {\bibfnamefont
  {D.}~\bibnamefont {Allwood}},\ }\href@noop {} {\bibfield  {journal} {\bibinfo
   {journal} {J. Appl. Phys.}\ }\textbf {\bibinfo {volume} {109}} (\bibinfo
  {year} {2011})}\BibitemShut {NoStop}%
\bibitem [{\citenamefont {Harres}\ and\ \citenamefont {Geshev}(2012)}]{harres}%
  \BibitemOpen
  \bibfield  {author} {\bibinfo {author} {\bibfnamefont {A.}~\bibnamefont
  {Harres}}\ and\ \bibinfo {author} {\bibfnamefont {J.}~\bibnamefont
  {Geshev}},\ }\href {http://stacks.iop.org/0953-8984/24/i=32/a=326004}
  {\bibfield  {journal} {\bibinfo  {journal} {J. Physics.: Condens. Matter}\
  }\textbf {\bibinfo {volume} {24}},\ \bibinfo {pages} {326004} (\bibinfo
  {year} {2012})}\BibitemShut {NoStop}%
\bibitem [{\citenamefont {Dubowik}\ \emph
  {et~al.}(2013{\natexlab{a}})\citenamefont {Dubowik}, \citenamefont
  {Go\'{s}cia\'{n}ska}, \citenamefont {Za{\l}\c{e}ski}, \citenamefont
  {G{\l}owi\'{n}ski}, \citenamefont {Kudryavtsev},\ and\ \citenamefont
  {Ehresmann}}]{dubowik}%
  \BibitemOpen
  \bibfield  {author} {\bibinfo {author} {\bibfnamefont {J.}~\bibnamefont
  {Dubowik}}, \bibinfo {author} {\bibfnamefont {I.}~\bibnamefont
  {Go\'{s}cia\'{n}ska}}, \bibinfo {author} {\bibfnamefont {K.}~\bibnamefont
  {Za{\l}\c{e}ski}}, \bibinfo {author} {\bibfnamefont {H.}~\bibnamefont
  {G{\l}owi\'{n}ski}}, \bibinfo {author} {\bibfnamefont {Y.}~\bibnamefont
  {Kudryavtsev}}, \ and\ \bibinfo {author} {\bibfnamefont {A.}~\bibnamefont
  {Ehresmann}},\ }\href@noop {} {\bibfield  {journal} {\bibinfo  {journal} {{J.
  Appl. Phys.}}\ }\textbf {\bibinfo {volume} {{113}}},\ \bibinfo {pages}
  {{193907}} (\bibinfo {year} {{2013}}{\natexlab{a}})}\BibitemShut {NoStop}%
\bibitem [{\citenamefont {Skomski}(2003)}]{skomski}%
  \BibitemOpen
  \bibfield  {author} {\bibinfo {author} {\bibfnamefont {R.}~\bibnamefont
  {Skomski}},\ }\href@noop {} {\bibfield  {journal} {\bibinfo  {journal} {J.
  Phys.: Condens. Matter}\ }\textbf {\bibinfo {volume} {15}},\ \bibinfo {pages}
  {R841} (\bibinfo {year} {2003})}\BibitemShut {NoStop}%
\bibitem [{\citenamefont {O'Handley}(1999)}]{ohandley}%
  \BibitemOpen
  \bibfield  {author} {\bibinfo {author} {\bibfnamefont {R.}~\bibnamefont
  {O'Handley}},\ }\href {http://books.google.pl/books?id=RKV1QgAACAAJ} {\emph
  {\bibinfo {title} {Modern Magnetic Materials: Principles and Applications}}}\
  (\bibinfo  {publisher} {Wiley},\ \bibinfo {year} {1999})\BibitemShut
  {NoStop}%
\bibitem [{\citenamefont {Frait}\ and\ \citenamefont {Fraitova}(1988)}]{frait}%
  \BibitemOpen
  \bibfield  {author} {\bibinfo {author} {\bibfnamefont {Z.}~\bibnamefont
  {Frait}}\ and\ \bibinfo {author} {\bibfnamefont {D.}~\bibnamefont
  {Fraitova}},\ }in\ \href@noop {} {\emph {\bibinfo {booktitle} {Modern
  problems in condensed matter sciences}}},\ \bibinfo {series and number} {Spin
  waves and magnetic excitations, Vol. 22.2},\ \bibinfo {editor} {edited by\
  \bibinfo {editor} {\bibfnamefont {A.~S.}\ \bibnamefont {Borovik-Romanov}}\
  and\ \bibinfo {editor} {\bibfnamefont {S.~K.}\ \bibnamefont {Sinha}}}\
  (\bibinfo  {publisher} {North-Holland},\ \bibinfo {address} {Amsterdam},\
  \bibinfo {year} {{1988}})\ \bibinfo {type} {Section}~\bibinfo {chapter} {1},
  pp.\ \bibinfo {pages} {1--65}\BibitemShut {NoStop}%
\bibitem [{\citenamefont {Liu}\ \emph {et~al.}(1994)\citenamefont {Liu},
  \citenamefont {Sooryakumar}, \citenamefont {Gutierrez},\ and\ \citenamefont
  {Prinz}}]{liu}%
  \BibitemOpen
  \bibfield  {author} {\bibinfo {author} {\bibfnamefont {X.}~\bibnamefont
  {Liu}}, \bibinfo {author} {\bibfnamefont {R.}~\bibnamefont {Sooryakumar}},
  \bibinfo {author} {\bibfnamefont {C.~J.}\ \bibnamefont {Gutierrez}}, \ and\
  \bibinfo {author} {\bibfnamefont {G.~A.}\ \bibnamefont {Prinz}},\ }\href
  {\doibase 10.1063/1.356763} {\bibfield  {journal} {\bibinfo  {journal} {J.
  Appl. Phys.}\ }\textbf {\bibinfo {volume} {75}},\ \bibinfo {pages} {7021}
  (\bibinfo {year} {1994})}\BibitemShut {NoStop}%
\bibitem [{\citenamefont {Trudel}\ \emph {et~al.}(2010)\citenamefont {Trudel},
  \citenamefont {Gaier}, \citenamefont {Hamrle},\ and\ \citenamefont
  {Hillebrands}}]{gaier}%
  \BibitemOpen
  \bibfield  {author} {\bibinfo {author} {\bibfnamefont {S.}~\bibnamefont
  {Trudel}}, \bibinfo {author} {\bibfnamefont {O.}~\bibnamefont {Gaier}},
  \bibinfo {author} {\bibfnamefont {J.}~\bibnamefont {Hamrle}}, \ and\ \bibinfo
  {author} {\bibfnamefont {B.}~\bibnamefont {Hillebrands}},\ }\href
  {{http://stacks.iop.org/0022-3727/43/i=19/a=193001}} {\bibfield  {journal}
  {\bibinfo  {journal} {{J. Phys. D: Applied Physics}}\ }\textbf {\bibinfo
  {volume} {{43}}},\ \bibinfo {pages} {{193001}} (\bibinfo {year}
  {{2010}})}\BibitemShut {NoStop}%
\bibitem [{\citenamefont {{R. Yilgin, S. Kazan, B. Rameev, B. Aktas, and K.
  Westerholt}}(2009)}]{yilgin}%
  \BibitemOpen
  \bibfield  {author} {\bibinfo {author} {\bibnamefont {{R. Yilgin, S. Kazan,
  B. Rameev, B. Aktas, and K. Westerholt}}},\ }\href {\doibase
  {10.1088/17426596/1/012068}} {\bibfield  {journal} {\bibinfo  {journal} {{J.
  Phys.: Conf. Series}}\ }\textbf {\bibinfo {volume} {{153}}},\ \bibinfo
  {pages} {{012068}} (\bibinfo {year} {{2009}})}\BibitemShut {NoStop}%
\bibitem [{\citenamefont {Dubowik}\ \emph
  {et~al.}(2013{\natexlab{b}})\citenamefont {Dubowik}, \citenamefont
  {Go{\'s}cia{\'n}ska}, \citenamefont {Za{\l}{\c e}ski}, \citenamefont
  {G{\l}owi{\'n}ski}, \citenamefont {Ehresmann}, \citenamefont {Kakazei},\ and\
  \citenamefont {Bunayev}}]{dubowik1}%
  \BibitemOpen
  \bibfield  {author} {\bibinfo {author} {\bibfnamefont {J.}~\bibnamefont
  {Dubowik}}, \bibinfo {author} {\bibfnamefont {I.}~\bibnamefont
  {Go{\'s}cia{\'n}ska}}, \bibinfo {author} {\bibfnamefont {K.}~\bibnamefont
  {Za{\l}{\c e}ski}}, \bibinfo {author} {\bibfnamefont {H.}~\bibnamefont
  {G{\l}owi{\'n}ski}}, \bibinfo {author} {\bibfnamefont {A.}~\bibnamefont
  {Ehresmann}}, \bibinfo {author} {\bibfnamefont {G.}~\bibnamefont {Kakazei}},
  \ and\ \bibinfo {author} {\bibfnamefont {S.~A.}\ \bibnamefont {Bunayev}},\
  }\href@noop {} {\bibfield  {journal} {\bibinfo  {journal} {{Acta Phys. Polon.
  A}}\ }\textbf {\bibinfo {volume} {{121}}},\ \bibinfo {pages} {{1121}}
  (\bibinfo {year} {{2013}}{\natexlab{b}})}\BibitemShut {NoStop}%
\bibitem [{\citenamefont {van~der Zaag}\ \emph {et~al.}(1996)\citenamefont
  {van~der Zaag}, \citenamefont {Ball}, , \citenamefont {Feiner}, \citenamefont
  {Wolf},\ and\ \citenamefont {van~der Heijden}}]{zaag}%
  \BibitemOpen
  \bibfield  {author} {\bibinfo {author} {\bibfnamefont {P.~J.}\ \bibnamefont
  {van~der Zaag}}, \bibinfo {author} {\bibfnamefont {A.~R.}\ \bibnamefont
  {Ball}}, , \bibinfo {author} {\bibfnamefont {L.~F.}\ \bibnamefont {Feiner}},
  \bibinfo {author} {\bibfnamefont {R.~M.}\ \bibnamefont {Wolf}}, \ and\
  \bibinfo {author} {\bibfnamefont {P.~A.~A.}\ \bibnamefont {van~der
  Heijden}},\ }\href@noop {} {\bibfield  {journal} {\bibinfo  {journal} {{J.
  Appl. Phys.}}\ }\textbf {\bibinfo {volume} {79}},\ \bibinfo {pages} {5103}
  (\bibinfo {year} {1996})}\BibitemShut {NoStop}%
\bibitem [{\citenamefont {Coehoorn}(2003)}]{coehoorn}%
  \BibitemOpen
  \bibfield  {author} {\bibinfo {author} {\bibfnamefont {R.}~\bibnamefont
  {Coehoorn}},\ }\href@noop {} {\enquote {\bibinfo {title} {Lecture notes on
  novel mag\-neto\-electro\-nic mater\-ials and devices: exchange biased
  spinvalves, part 4.}}\ }\bibinfo {howpublished}
  {{http://www.tue.nl/en/university/departments/applied-physics/research/funct%
ional-materials/physics-of-nanostructures/students-education/lectures-courses/%
}} (\bibinfo {year} {2003}),\ \bibinfo {note} {[Online; accessed
  19-July-2013]}\BibitemShut {NoStop}%
\bibitem [{\citenamefont {Tsunoda}\ \emph {et~al.}(2006)\citenamefont
  {Tsunoda}, \citenamefont {Imakita}, \citenamefont {Naka},\ and\ \citenamefont
  {Takahashi}}]{tsunoda}%
  \BibitemOpen
  \bibfield  {author} {\bibinfo {author} {\bibfnamefont {M.}~\bibnamefont
  {Tsunoda}}, \bibinfo {author} {\bibfnamefont {K.}~\bibnamefont {Imakita}},
  \bibinfo {author} {\bibfnamefont {M.}~\bibnamefont {Naka}}, \ and\ \bibinfo
  {author} {\bibfnamefont {M.}~\bibnamefont {Takahashi}},\ }\href@noop {}
  {\bibfield  {journal} {\bibinfo  {journal} {J. Magn. Magn. Mater.}\ }\textbf
  {\bibinfo {volume} {304}},\ \bibinfo {pages} {55 } (\bibinfo {year}
  {2006})}\BibitemShut {NoStop}%
\bibitem [{\citenamefont {Aharoni}\ \emph {et~al.}(1959)\citenamefont
  {Aharoni}, \citenamefont {Frei},\ and\ \citenamefont {Shtrikman}}]{aharoni}%
  \BibitemOpen
  \bibfield  {author} {\bibinfo {author} {\bibfnamefont {A.}~\bibnamefont
  {Aharoni}}, \bibinfo {author} {\bibfnamefont {E.~H.}\ \bibnamefont {Frei}}, \
  and\ \bibinfo {author} {\bibfnamefont {S.}~\bibnamefont {Shtrikman}},\
  }\href@noop {} {\bibfield  {journal} {\bibinfo  {journal} {J. Appl. Phys.}\
  }\textbf {\bibinfo {volume} {30}},\ \bibinfo {pages} {1956} (\bibinfo {year}
  {1959})}\BibitemShut {NoStop}%
\bibitem [{\citenamefont {Salansky}\ and\ \citenamefont
  {Eruchimov}(1970)}]{salansky}%
  \BibitemOpen
  \bibfield  {author} {\bibinfo {author} {\bibfnamefont {N.~M.}\ \bibnamefont
  {Salansky}}\ and\ \bibinfo {author} {\bibfnamefont {M.~S.}\ \bibnamefont
  {Eruchimov}},\ }\href@noop {} {\bibfield  {journal} {\bibinfo  {journal}
  {{Thin Solid Films}}\ }\textbf {\bibinfo {volume} {{6}}},\ \bibinfo {pages}
  {{129}} (\bibinfo {year} {{1970}})}\BibitemShut {NoStop}%
\bibitem [{\citenamefont {van~der Zaag}\ and\ \citenamefont
  {Brochers}(2010)}]{zaagbook}%
  \BibitemOpen
  \bibfield  {author} {\bibinfo {author} {\bibfnamefont {P.~J.}\ \bibnamefont
  {van~der Zaag}}\ and\ \bibinfo {author} {\bibfnamefont {J.~A.}\ \bibnamefont
  {Brochers}},\ }in\ \href {http://dnb.d-nb.de} {\emph {\bibinfo {booktitle}
  {Magnetic properties of antiferromagnetic oxide materials}}},\ \bibinfo
  {editor} {edited by\ \bibinfo {editor} {\bibfnamefont {L.}~\bibnamefont
  {D\'{u}o}}, \bibinfo {editor} {\bibfnamefont {M.}~\bibnamefont {Finazzi}}, \
  and\ \bibinfo {editor} {\bibfnamefont {F.}~\bibnamefont {Ciccacci}}}\
  (\bibinfo  {publisher} {Wiley-VCH},\ \bibinfo {year} {2010})\ Chap.~\bibinfo
  {chapter} {7}\BibitemShut {NoStop}%
\bibitem [{\citenamefont {Ball}\ \emph {et~al.}(1996)\citenamefont {Ball},
  \citenamefont {Leenaers}, \citenamefont {van~der Zaag}, \citenamefont {Shaw},
  \citenamefont {Singer}, \citenamefont {Lind}, \citenamefont {Fredrikze},\
  and\ \citenamefont {Rekveldt}}]{ball}%
  \BibitemOpen
  \bibfield  {author} {\bibinfo {author} {\bibfnamefont {A.~R.}\ \bibnamefont
  {Ball}}, \bibinfo {author} {\bibfnamefont {A.~J.~G.}\ \bibnamefont
  {Leenaers}}, \bibinfo {author} {\bibfnamefont {P.~J.}\ \bibnamefont {van~der
  Zaag}}, \bibinfo {author} {\bibfnamefont {K.~A.}\ \bibnamefont {Shaw}},
  \bibinfo {author} {\bibfnamefont {B.}~\bibnamefont {Singer}}, \bibinfo
  {author} {\bibfnamefont {D.~M.}\ \bibnamefont {Lind}}, \bibinfo {author}
  {\bibfnamefont {H.}~\bibnamefont {Fredrikze}}, \ and\ \bibinfo {author}
  {\bibfnamefont {M.}~\bibnamefont {Rekveldt}},\ }\href@noop {} {\bibfield
  {journal} {\bibinfo  {journal} {Appl. Phys. Lett.}\ }\textbf {\bibinfo
  {volume} {69}},\ \bibinfo {pages} {1489} (\bibinfo {year}
  {1996})}\BibitemShut {NoStop}%
\bibitem [{\citenamefont {Radu}\ and\ \citenamefont {Zabel}(2008)}]{radu}%
  \BibitemOpen
  \bibfield  {author} {\bibinfo {author} {\bibfnamefont {F.}~\bibnamefont
  {Radu}}\ and\ \bibinfo {author} {\bibfnamefont {H.}~\bibnamefont {Zabel}},\
  }in\ \href {\doibase 10.1007/978-3-540-73462-8_3} {\emph {\bibinfo
  {booktitle} {Magnetic Heterostructures}}},\ \bibinfo {series} {Springer
  Tracts in Modern Physics}, Vol.\ \bibinfo {volume} {227},\ \bibinfo {editor}
  {edited by\ \bibinfo {editor} {\bibfnamefont {H.}~\bibnamefont {Zabel}}\ and\
  \bibinfo {editor} {\bibfnamefont {S.~D.}\ \bibnamefont {Bader}}}\ (\bibinfo
  {publisher} {Springer Berlin Heidelberg},\ \bibinfo {year} {2008})\ pp.\
  \bibinfo {pages} {97--184}\BibitemShut {NoStop}%
\bibitem [{\citenamefont {{H. Endo, A. Hirohata, T. Nakayama, and K.
  O'Grady}}(2011)}]{endo}%
  \BibitemOpen
  \bibfield  {author} {\bibinfo {author} {\bibnamefont {{H. Endo, A. Hirohata,
  T. Nakayama, and K. O'Grady}}},\ }\href@noop {} {\bibfield  {journal}
  {\bibinfo  {journal} {{J. Phys. D: Applied Physics}}\ }\textbf {\bibinfo
  {volume} {{44}}},\ \bibinfo {pages} {{14003}} (\bibinfo {year}
  {{2011}})}\BibitemShut {NoStop}%
\bibitem [{\citenamefont {O'Grady}\ \emph {et~al.}(2010)\citenamefont
  {O'Grady}, \citenamefont {Fernandez-Outon},\ and\ \citenamefont
  {Vallejo-Fernandez}}]{ogrady}%
  \BibitemOpen
  \bibfield  {author} {\bibinfo {author} {\bibfnamefont {K.}~\bibnamefont
  {O'Grady}}, \bibinfo {author} {\bibfnamefont {L.}~\bibnamefont
  {Fernandez-Outon}}, \ and\ \bibinfo {author} {\bibfnamefont {G.}~\bibnamefont
  {Vallejo-Fernandez}},\ }\href {\doibase
  http://dx.doi.org/10.1016/j.jmmm.2009.12.011} {\bibfield  {journal} {\bibinfo
   {journal} {J. Magn. Magn. Mater.}\ }\textbf {\bibinfo {volume} {322}},\
  \bibinfo {pages} {883 } (\bibinfo {year} {2010})}\BibitemShut {NoStop}%
\end{thebibliography}%

\end{document}